\documentclass[%
 reprint,
superscriptaddress,
 amsmath,amssymb,
 aps,
]{revtex4-1}

\usepackage{graphicx}
\usepackage{dcolumn}
\usepackage{bm}
\usepackage{braket}
\usepackage{color}
\usepackage{latexsym}
\usepackage{amsmath,amsthm}
\usepackage{amsfonts}
\usepackage{amsmath,amssymb,amsthm,mathrsfs,amsfonts,dsfont}
\usepackage{subfigure, epsfig}
\usepackage{bm}
\usepackage{enumerate}
\usepackage{color}
\usepackage{graphicx}
\usepackage{mathtools}
\usepackage{times}
\usepackage{appendix}
\usepackage{physics}

\newcommand{\YM}[1]{\textcolor[rgb]{0.1, 0.5, 0.1}{#1}}

\newcommand{\ma}[1]{\left({#1}\right)}
\newcommand{\na}[1]{\left\{{#1}\right\}}
\newcommand{\ka}[1]{\left[{#1}\right]}

\begin{document}
\allowdisplaybreaks
\jot=1pt


\title{
Quantum battery based on superabsorption
}



 \author{Yudai Ueki}
 \affiliation{Faculty of Pure and Applied Sciences, University of Tsukuba, Tsukuba 305-8571, Japan}

 \author{Shunsuke Kamimura}
 \affiliation{Faculty of Pure and Applied Sciences, University of Tsukuba, Tsukuba 305-8571, Japan}
 \affiliation{%
 Research Center for Emerging Computing Technologies, National Institute of Advanced Industrial Science and Technology (AIST),
1-1-1 Umezono, Tsukuba, Ibaraki 305-8568, Japan.
}%
 \author{Yuichiro Matsuzaki}
 \email{matsuzaki.yuichiro@aist.go.jp}
\affiliation{%
 Research Center for Emerging Computing Technologies, National Institute of Advanced Industrial Science and Technology (AIST),
1-1-1 Umezono, Tsukuba, Ibaraki 305-8568, Japan.
}%
\affiliation{
NEC-AIST Quantum Technology Cooperative Research Laboratory,
National Institute of Advanced Industrial Science and Technology (AIST), Tsukuba, Ibaraki 305-8568, Japan
}

\author{Kyo Yoshida}
\affiliation{Faculty of Pure and Applied Sciences, University of Tsukuba, Tsukuba 305-8571, Japan}
\author{Yasuhiro Tokura 
}%
\email{tokura.yasuhiro.ft@u.tsukuba.ac.jp}
\affiliation{Faculty of Pure and Applied Sciences, University of Tsukuba, Tsukuba 305-8571, Japan}
 \email{tokura.yasuhiro.ft@u.tsukuba.ac.jp}

\date{\today}

\begin{abstract}
A quantum battery is a device where an energy is charged by using a quantum effect.
Here, we propose a quantum battery with a charger system composed of $N$ qubits by
utilizing a collective effect called a superabsorption. Importantly, the coupling strength between the quantum battery and charger system can be enhanced due to an entanglement.
While the charger time scales as $\Theta\ma{N^{-1/2}}$ by applying a conventional scheme,
we can achieve a charging time $\Theta\ma{N^{-1}}$ in our scheme. 
Our results open the path to ultra-fast charging of a quantum battery.

\end{abstract}

\maketitle

\section{Introduction}
Quantum thermodynamics
is an emerging field to extend the conventional thermodynamics to microscopic systems where not only thermal but
also quantum fluctuations should be taken into account \cite{gemmer2009quantum,horodecki2013fundamental,pekola2015towards,goold2016role,vinjanampathy2016quantum,binder2018thermodynamics,deffner2019quantum}. 
One of the aims of quantum thermodynamics is to investigate whether quantum devices provide enhancement of performance over classical devices.

A quantum heat engine is one of the promising devices with the enhancement over the classical ones by using the quantum property \cite{scovil1959three,quan2007quantum,funo2013thermodynamic,kosloff2014quantum,rossnagel2014nanoscale,tajima2021superconducting,kloc2021superradiant,kamimura2021quantum,souza2021collective}.
It is possible to obtain a
quadratically scaling
power $P=\Theta (N^2)$ by using the quantum enhanced heat engine with an entanglement
while a conventional separable engine shows a power of $P=\Theta (N)$ 
where $N$ denotes the number of qubits \cite{kamimura2021quantum}.
Here, the key feature of this scheme is to adopt a collective effect called a superabsorption that was proposed in Ref.~\cite{higgins2014superabsorption},
and its proof-of-concept experiment
was recently demonstrated
using barium atoms in an optical cavity
\cite{yang2021realization}.

A quantum battery is also a prominent research subject in quantum thermodynamics
to charge an energy of quantum systems.
As is the case for conventional batteries (such as lithium-ion batteries) using electrochemical reactions to convert chemical energy into electrical energy,
the main issue of the quantum battery is to increase a performance of charging and discharging processes \cite{alicki2013entanglement,campaioli2017enhancing,campaioli2018quantum,ferraro2018high}.
Such a quantum battery was first proposed in Ref.~\cite{alicki2013entanglement}.
In Ref.~\cite{campaioli2017enhancing}, it was found that
the use of entangling operations can improve the performance of the quantum battery compared with the one
using only separable operations.
Here, the performance is defined as a 
storable energy in a quantum battery per a unit time.
Also, in Ref.~\cite{ferraro2018high}, it has been found that Dicke-quantum batteries composed of collective $N$-qubit systems give us a scaling $\sqrt{N}$ times larger
compared to independent $N$-qubit batteries.
In these studies, external pulses are applied
to charge the isolated quantum battery.

On the other hand, the battery can also be charged by using an interaction with an environment \cite{tacchino2020charging,ito2020collectively}.
In Ref. \cite{ito2020collectively}, when $N$ quantum batteries interact with $N$ charger systems that are prepared in a steady state with a population inversion, the charging time scales as $\Theta(N)$ by using a collective charging process.
Here, the charging time is defined as how long it takes for the quantum battery to be a steady state.
In Ref.~\cite{quach2022superabsorption}, the improvement of the quantum battery performance due to the collective effects has been experimentally confirmed, and the charging time scales as $\Theta\ma{N^{-1/2}}$ where $N$ denotes the number of atoms.

Here, we propose a quantum battery using a charger system composed of an entangled $N$-qubit system.
Our quantum battery provides a charging time scaling as $\Theta(N^{-1})$.
This is stark contrast to the conventional scheme where the charging time scales as $\Theta (N^{-1/2})$ by using $N$ three-level systems as a charger.
The key factor for the enhanced charging time is utilizing the superabsorption for the charging process.
The charger systems are prepared in an
entangled $N$-qubit state called
Dicke state via
an interaction with the environment, and the quantum battery can strongly interact with the charger system in a collective way.
An energy exchange between the charger
and the
battery occurs, and the battery, initially prepared in a ground state, can be eventually raised
in an excited state
with a necessary time scaling as $\Theta (N^{-1})$.

Our paper is organized as follows. In Sec.~II, we review a charging model with one three-level system and one qubit battery. Also, we explain a charging model using
$N$-qubit charger initially prepared in a separable state,
where the charging time scales as $\Theta (N^{-1/2})$. In Sec. III, we explain our charging scheme
with a charging time scaling as $\Theta (N^{-1})$ by using an $N$-qubit charger initially prepared in an entangled state.
In Sec. IV, we conclude our discussion.

\section{Battery charging with separable N three-level chargers}
Let us
review
the conventional
charging model with separable states.
We consider the charger system with a three-level system
and the quantum battery system with a qubit. The total Hamiltonian $H_{tot}^{1}$ is given as follows.
\begin{align}
H_{tot}^1&=H_{\mathrm{3-level}}+H_B^1+H_I^1,
\\
H_{\mathrm{3-level}}&=\sum_{i=0}^2 E_i \ket{i}_C\bra{i},
\\
H_B^1&=\frac{\Delta}{2}\sigma_z,
\ \ \Delta\equiv E_1-E_0,
\\
H_I^1&=g\ma{
\ket{0}_C\bra{1}\otimes \ket{1}_B\bra{0}
+\mathrm{h.c.}
},
\end{align}
where $E_i$ ($i=0,1,2, E_0<E_1<E_2$)
is the eigenenergy of the
charger system,
$\Delta$
is the energy of the
quantum battery system,
$g$ is a coupling strength between the
charger and quantum
battery. 
We prepare the initial system $\ket{\psi^1(0)}=\ket{1}_C\ket{0}_B$ where the charger state is prepared as $\ket{1}_C$ and the battery state is prepared as $\ket{0}_B$. 
A steady state of the charger system after a 
coupling with two thermal baths can be $\ket{1}_C$ by adjusting the parameters where one of the thermal baths induces a transition between $\ket{0}_C$ and $\ket{2}_C$ while the other thermal baths
induces a transition between $\ket{2}_C$ and $\ket{1}_C$, and this is called a population inversion
where the population of the first excited state becomes higher than that of the ground state
\cite{higgins2014superabsorption,ito2020collectively,kloc2021superradiant}.
The purpose of the charging is to obtain a battery state of $\ket{1}_B$ from the initial state. The battery state $\rho^1_B(t)$ can be described as 
\begin{align}
\rho^1_B(t)&=
\mathrm{Tr}_C\ka{e^{-iH^1_{tot}t}\ket{\psi^1(0)}\bra{\psi^1(0)}
e^{iH^1_{tot}t}
}
\nonumber
\\
&=
\cos^2(gt)\ket{0}_B\bra{0}+\sin^2(gt)\ket{1}_B\bra{1},
\end{align}
where $\mathrm{Tr}_C$ denotes partial trace of the charger system. When we turn off the interaction at $gt=\pi/2$, we obtain a state of$\ket{1}_B$, and this means that the charging is done.
\begin{figure*}[tbp]
 \begin{minipage}{0.33\hsize}
  \begin{center}
   \includegraphics[width=55mm]{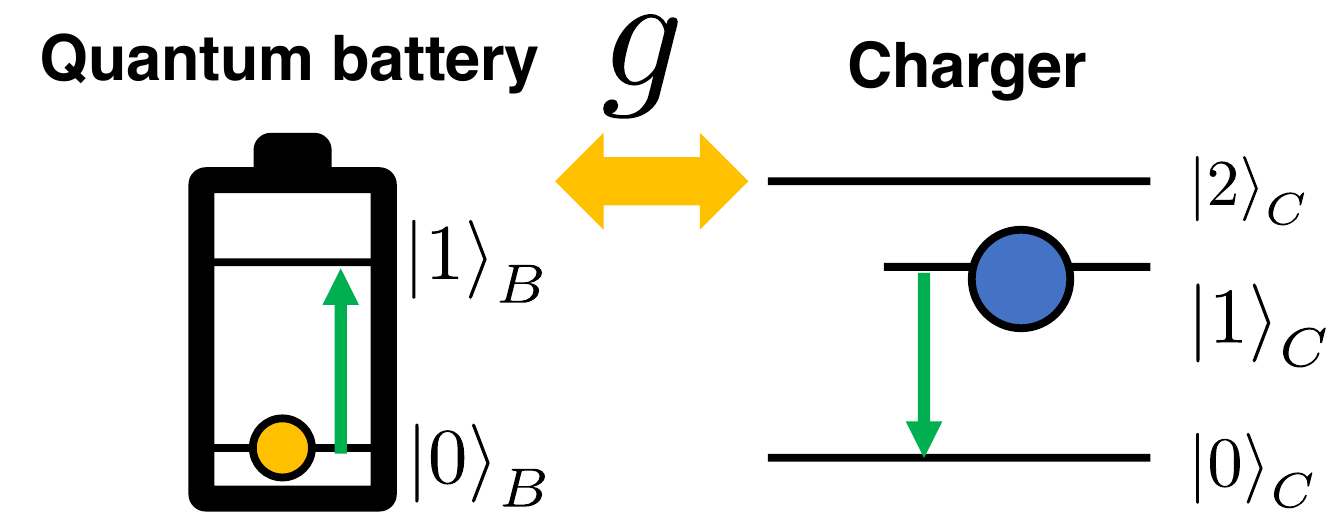}
   \caption{Schematic of three-level charging model.}
   \label{fig:3level_charging}
  \end{center}
 \end{minipage}
 
 \begin{minipage}{0.5\hsize}
 \begin{center}
  \includegraphics[width=80mm]{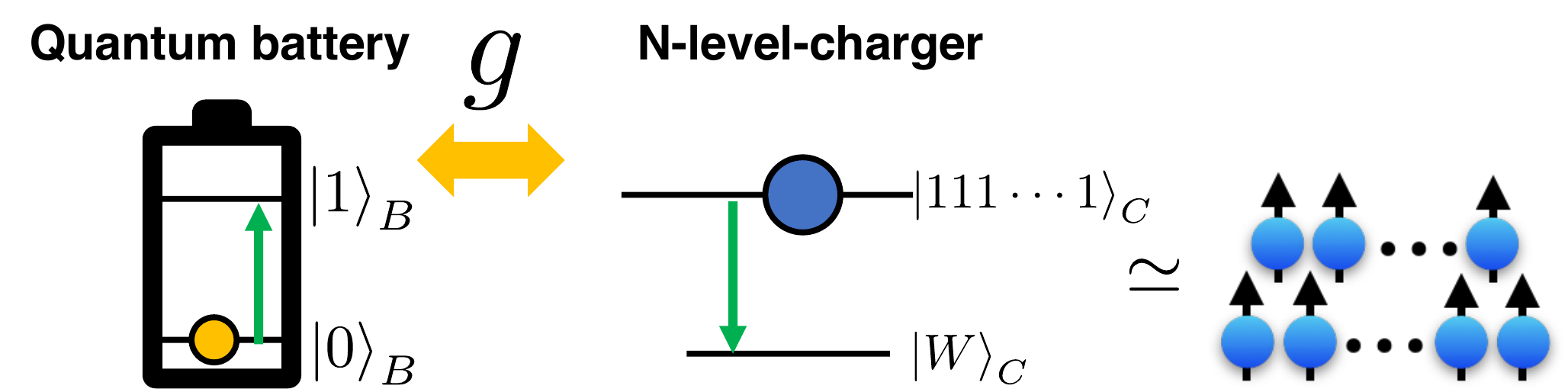}
   \caption{Schematic of N-level charging model.Each frequency of qubits are $\omega_A$ and initial state of qubits are excited state $\ket{1}_C$ .}
   \label{fig:Nlevel_charging}
 \end{center}
 \end{minipage}
\end{figure*}

Next, let us explain
a scheme to use $N$ three-level systems as a charger.
We consider the charger system with $N$ three-level systems and the quantum battery system with a qubit. 
Strictly speaking, we need three-level systems for the charger systems, because this allows us to use a population inversion when the charger system becomes a steady state after the coupling with thermal baths, as we mentioned above. 
However, once we successfully obtain the population inversion for the charger system, the dynamics between the quantum battery and charger system is confined in a subspace spanned by $\ket{0}_C$ and $\ket{1}_C$, and so we consider only this subspace for simplicity.

We define the collective operators $J_z$ and $J_\pm$
as 
$J_z=\frac{1}{2}\sum_{i=1}^N \sigma^{i}_z$ and $J_{\pm}=\sum_{i=1}^{N} \sigma_{\pm}^i$.
The total Hamiltonian $H^{\mathrm{sep}}_{tot}$ is given by follows.
\begin{align}
H^{\mathrm{sep}}_{tot}
&= 
H^{\mathrm{sep}}_N+H^{\mathrm{sep}}_B+H^{\mathrm{sep}}_I,
\\
H_N&=\omega_AJ_z,
\\
H^{\mathrm{sep}}_B&=\frac{\omega_A}{2}\sigma_z,
\\
H^{\mathrm{sep}}_I
&=g\ma{\sigma_+\otimes J_{-}+\sigma_-\otimes J_{+}}
,
\end{align}
where $\omega_A$ denotes a frequency of the qubits for the charger system and a quantum battery and  $g$ is coupling strength between the charger and quantum battery. 
This Hamiltonian was experimentally realized by using a superconducting qubit and an electron-spin ensemble \cite{kubo2011hybrid,zhu2011coherent,saito2013towards,matsuzaki2015improving}.
We prepare the initial system $\ket{\psi^N(0)}=\ket{11\cdots 1}_C\ket{0}_B$ where the charger state is prepared as all excited states, $\ket{11\cdots 1}_C$, and the battery state is prepared as $\ket{0}_B$. 
The purpose of the charging is to obtain a battery state of $\ket{1}_B$ from the initial state. The battery state $\rho^N_B(t)$ can be described as 
\begin{align}
&\rho^N_B(t)
\nonumber
\\
&=
\mathrm{Tr}_C\ka{
e^{-iH_N^{\mathrm{sep}}t}\ket{\psi^N(0)}\bra{\psi^N(0)}
e^{iH^{\mathrm{sep}}_Nt}
}
\nonumber
\\
&=
\cos^2\ma{\sqrt{N}gt}
\ket{11\cdots 1}_C\ket{0}_B
\nonumber
\\
&
+
\sin^2\ma{\sqrt{N}gt}
\ket{W}_C\ket{1}_B,
\\
\ket{W}_C&=
\frac{1}{\sqrt{N}}
\ma{
\ket{111\cdots 0}_C+\cdots +
\ket{011\cdots 1}_C
}.
\end{align}
From this analysis
the necessary time to obtain $\ket{1}_B$ from $\ket{0}_B$ is $t=\pi /2\sqrt{N}g$.
This means that the charging time for the battery scale as $\Theta (N^{-1/2})$ in this model.
Such a behavior was theoretically predicted in 
\cite{marcos2010coupling,twamley2010superconducting}.

\section{Battery charging with superabsorption}
Here, we introduce our scheme to charge the quantum battery with a charging time to scale as $\Theta (N^{-1})$ by using $N$ charger qubits.

\subsection{Hamiltonian}
\begin{figure}[tbp]
\begin{center}
\small
\includegraphics[keepaspectratio, scale=0.6]{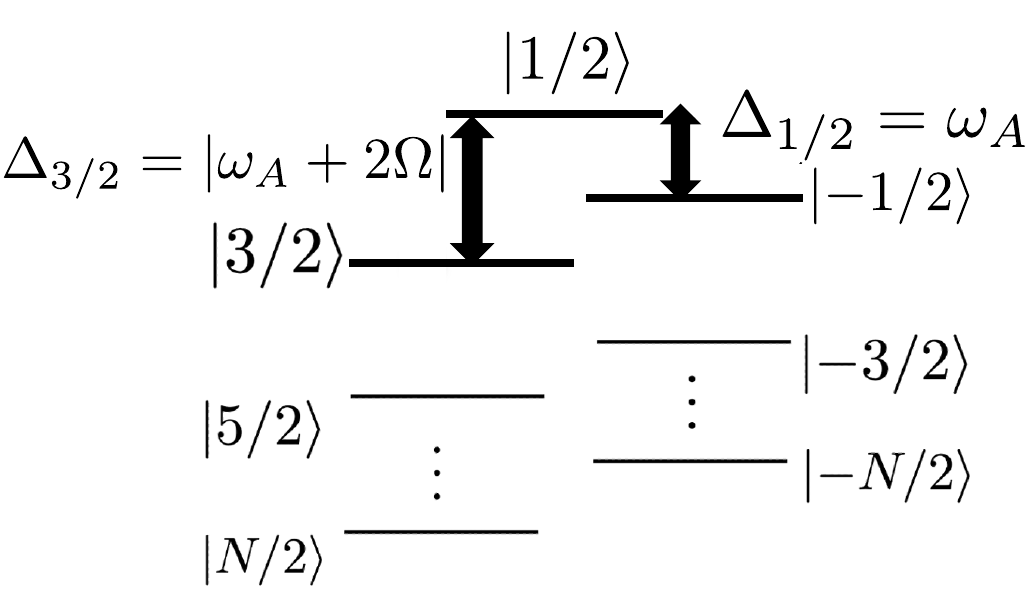}
\caption{Schematic of $\Lambda$-type Dicke states.}
\label{fig:Dicke_ladder}
\end{center}
\end{figure}

We consider the charger system and the quantum battery system. The former one is composed of $N$-qubits while the latter one is composed of two-qubits.
The total Hamiltonian $H_{tot}$ is given as follows.
\begin{align}
    H_{tot}&=H_N+H_B+H_{I},
    \\
  H_N&=\omega_AJ_z+\Omega J_z^2,
  \\
  H_B&=
  \frac{\omega_A+\delta}{2} \sigma_z^{(1)}
  +
  \frac{\abs{\omega_A+2\Omega}+\delta}{2} \sigma_z^{(2)},
  \\ 
  H_I&=
   2g\ma{
  \sigma_x^{(1)}+\sigma_x^{(2)}
  }\otimes J_x,
\end{align}
where $H_N(H_B)$ denotes the Hamiltonian for $N$-qubit system (2-qubit system), $H_I$ denotes the interaction Hamiltonian between charger ($N$-qubit) and quantum battery (2-qubit), $\omega_A$ denotes a frequency of the each $N$-qubits, $\Omega$ denotes a total coupling constant between the $N$-qubits, $g$ denotes a coupling constant between charger and battery. 
Here, let us introduce $J_{\pm 1}$ and $J_{\pm 2}$, which are the part of the ladder operator. They are defined as follows.
\begin{align}
      J_+&=J_{+1}+J_{+2}
  ,\ \
  J_-=J_{-1}+J_{-2},
  \\
  J_{+1}&=
 \sum_{M=3/2}^{N/2}\sqrt{a_{M}}\ket{M}_C\bra{M-1}
  , \ \  J_{-1}=(J_{+1})^{\dagger},
   \\
  J_{+2}&=\sum_{M=1-N/2}^{1/2}\sqrt{a_{M}}\ket{M}_C\bra{M-1}
  , \ \ J_{-2}=(J_{+2})^{\dagger},
  \\
  a_M&\equiv
\ma{
\frac{N}{2}+M
}\ma{
\frac{N}{2}-M+1
}.
\end{align}
Here, we introduced the Dicke states, which are the simultaneous eigenstates of $J^2$ and $J_z$. These can be written as $\ket{J,M}$, and the corresponding eigenvalues are $J(J+1)$ and $M$.
In this paper, we take Dicke states as $\ket{M}=\ket{N/2,M}$ in the subspace with total angular momentum $J=N/2$ and assume $N$ is odd.
We assume conditions of strong coupling, i.e.,
$\abs{\Omega}>\omega_A$. Also, we assume $\omega_A>0$ and $\Omega<0$.
Theses conditions allow us to construct a $\Lambda$-type structure for the
Dicke states between $\ket{3/2}_C$, $\ket{1/2}_C$, and $\ket{-1/2}_C$ as shown in the FIG \ref{fig:Dicke_ladder}.
In this case, $\ket{1/2}_C$ has the highest energy in the charger system.
This means that $J_{+1}$ and $J_{-2}$ play a role in inducing a transition from a higher energy state to a lower energy state in the charger system.
On the other hand, $J_{-1}$ and $J_{+2}$ induce a transition from a lower energy state to a higher energy state in the charger system.
We are going to use a rotating wave approximation (RWA) for $gN\ll \omega_A$.  
In the RWA, we typically ignore terms that oscillate with a high frequency. In our case, terms such as $(\sigma _-^{(1)} +\sigma _-^{(2)})J_{+1}$, $(\sigma _-^{(1)} +\sigma _-^{(2)})J_{-2}$, $(\sigma _+^{(1)} +\sigma _+^{(2)})J_{-1}$, $(\sigma _+^{(1)} +\sigma _+^{(2)})J_{+2}$ will be dropped.
So, by using the RWA, we obtain $H_I\simeq g(A+A^{\dagger})$ where $A\equiv \ma{\sigma_+^{(1)}+\sigma_+^{(2)}}(J_{+1}+J_{-2})$.

In this case, the Hamiltonian of the $N$-qubit system can be diagonalized as follows. 
\begin{align}
    H_N&=\sum_{M=-N/2}^{N/2} E_M \ket{M}_C\bra{M},\ \ 
    E_M=\omega_AM+\Omega M^2,
    \\
    M&\in \na{-\frac{N}{2},-\frac{N}{2}+1,\cdots,
    \frac{N}{2}}.
\end{align}
Also, we describe the energy eigenstates
of the $N$-qubit system as the Dicke states. 
The energy differences between the Dicke states are
written as $\Delta_M=E_M-E_{M-1}=\omega_A+\Omega(2M-1)$.
Let us denote the frequency of the qubit 1 by
$\omega_A+\delta$, which
is detuned from  $\Delta_{1/2}$ by
$\delta$.
On the other hand, $\abs{\omega_A+2\Omega}+\delta$
denotes the frequency of qubit 2, and this is detuned from $\Delta_{3/2}$ by $\delta$.

\subsection{Perturbation analysis}
Let us derive an effective Hamiltonian. We move into an interaction picture.
We consider a unitary operator
$U(t,0)
=\mathcal{T}e^{-i\int^t_0 H^i_I(t)dt}$
where $H^i_I(t)\equiv e^{i(H_N+H_B)t}H_I e^{-i(H_N+H_B)t}$ denotes an interaction Hamiltonian in the interaction picture. We expand
$U(t,0)$ to the second order term with respect to the coupling constant $g$.
\begin{eqnarray}
&&U(t,0)
=I-i\int^{t}_0 H^i_I(t^{\prime})dt^{\prime}
\nonumber 
\\
&&  -\int^{t}_0 H^i_I(t^{\prime})dt^{\prime}
  \int^{t^{\prime}}_0
  H^i_I(t^{\prime\prime})dt^{\prime\prime}
  +\mathcal{O}(g^3)\label{ymadd}.
\end{eqnarray}
Under the assumptions of
$gN\ll \delta \ll \omega_A < \abs{\Omega}$ and $1\ll gNt $, the first order of the interaction term induces transitions between states with a large energy difference, and these have terms to oscilalte with a high frequency.
So the second order term, which includes a resonant transition,
becomes the relevant term.
Then we obtain
\begin{align}
U(t,0)&=
I-itH_{\mathrm{eff}}\simeq e^{-itH_{\mathrm{eff}}}.
\\
H_{\mathrm{eff}}&=
\frac{g^2}{\delta}
  \sqrt{a_{1/2}}\sqrt{a_{3/2}}
  \ma{
  \tilde{L}_++\tilde{L}_-
  }\ .
  \\
  \tilde{L}_+&=
  \sigma^{(2)}_+\sigma^{(1)}_-\ket{3/2}_C\bra{-1/2}
  ,\ \tilde{L}_-=\ma{\tilde{L}_+}^{\dagger}\ .
  \nonumber 
\end{align}
Here, when the initial state of the battery is confined in a subspace spanned by $\na{\ket{01}_B, \ket{10}_B}$,
the dynamics of the battery state is also confined in this subspace as long as the second order term is relevant.
(See the appendix for the details of the derivation).
Although a similar approximation has been used in quantum optics
\cite{gerry2005introductory}, we firstly applied this technique to the model of the quantum battery with a superabsorption.

\begin{figure}[tbp]
\begin{center}
\small
\includegraphics[keepaspectratio, scale=0.45]{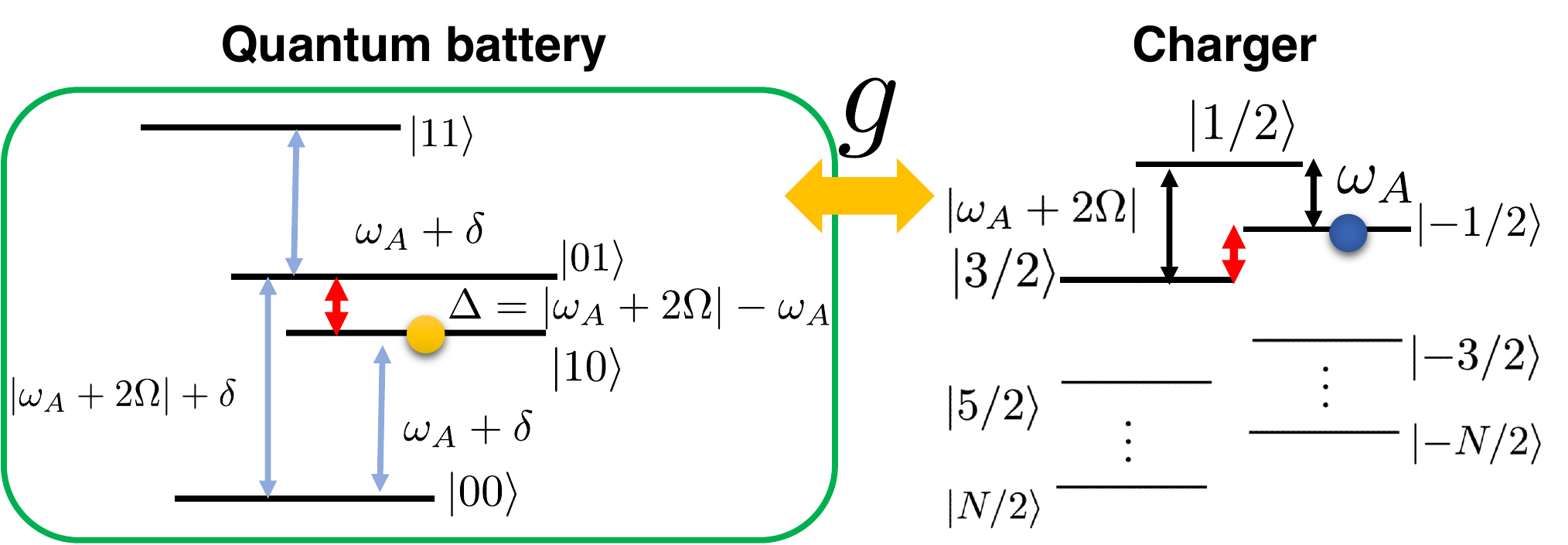}
\caption{Schematic of battery charging with our model.
The initial state of a battery is $\ket{10}_B$ and charger is $\ket{-1/2}_C$.
}
\label{fig:charging_process}
\end{center}
\end{figure}

We discuss the dynamics of the system with the effective Hamiltonian.
We prepare the initial state $\ket{\psi(0)}=\ket{-1/2}_C\ket{10}_B$ 
(see FIG \ref{fig:charging_process}).
The purpose of the charging is to obtain a battery state of $\ket{01}_B$ from the initial state.
Considering the charging process using the effective Hamiltonian $H_{\mathrm{eff}}$, the battery state $\rho^{\mathrm{eff}}_B(t)$ can be described as
\begin{align}
\rho^{\mathrm{eff}}_B(t)&=
\mathrm{Tr}_{N}\ka{
e^{-iH_{\mathrm{eff}}t}\ket{\psi(0)}\bra{\psi(0)}
e^{iH_{\mathrm{eff}}t}
}
\nonumber
\\
&=
\cos^2\ma{\lambda t}\ket{10}_C\langle 10|
+\sin^2\ma{\lambda t}\ket{01}_C\langle 01|
.
\end{align}
Here, $\lambda=\frac{g^2}{\delta}
  \sqrt{a_{1/2}}\sqrt{a_{3/2}}$ and 
$\mathrm{Tr}_N$ denotes a
partial trace of the charger system.
We set $\delta=10gN$ to satisfy condition $gN\ll \delta$.
Since $\sqrt{a_{1/2}}\sqrt{a_{3/2}}\propto \mathcal{O}(N^2)$ and $\delta \propto \mathcal{O}(N)$,
the necessary time to obtain $\ket{01}_B$ from the initial state of $\ket{10}_B$ scales as $\mathcal{O}(N^{-1})$. This means that the charging time for the battery scales as $\mathcal{O}(N^{-1})$.
We use a phenomena called super-absorption where the coupling strength becomes collectively enhanced around the middle of the Dicke ladder structure.
However, we use approximations to derive the effective Hamiltonian. To check the validity of the approximation, we will perform numerical simulations in the next subsection.

\subsection{Numerical analysis}
\begin{figure}[tbp]
\begin{center}
\small
\includegraphics[keepaspectratio, scale=0.5]{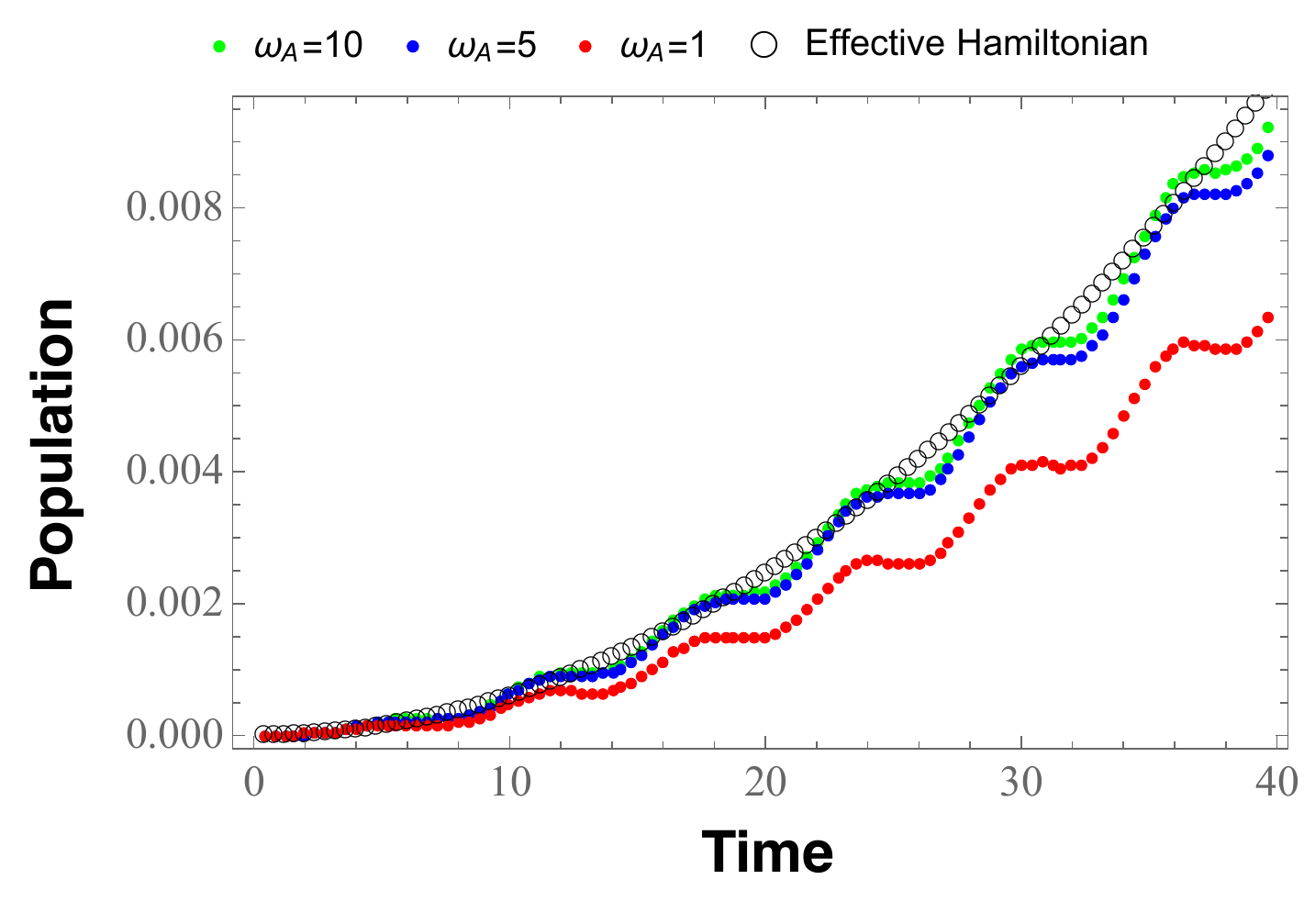}
\caption{
Plot of the population of $\ket{01}_B$ when we use the 
effective Hamiltonian or the exact Hamiltonian with $\omega_A=10,\ 5,\ 1$. 
We set the other parameters as
$g=10^{-3}, \ \delta=10gN,\ \Omega=-2.3\omega_A$ and $N=101$.
}
\label{fig:Compare_omegaA}
\end{center}
\end{figure}

\begin{figure*}[tbp]
 \begin{minipage}{0.33\hsize}
  \begin{center}
   \includegraphics[width=55mm]{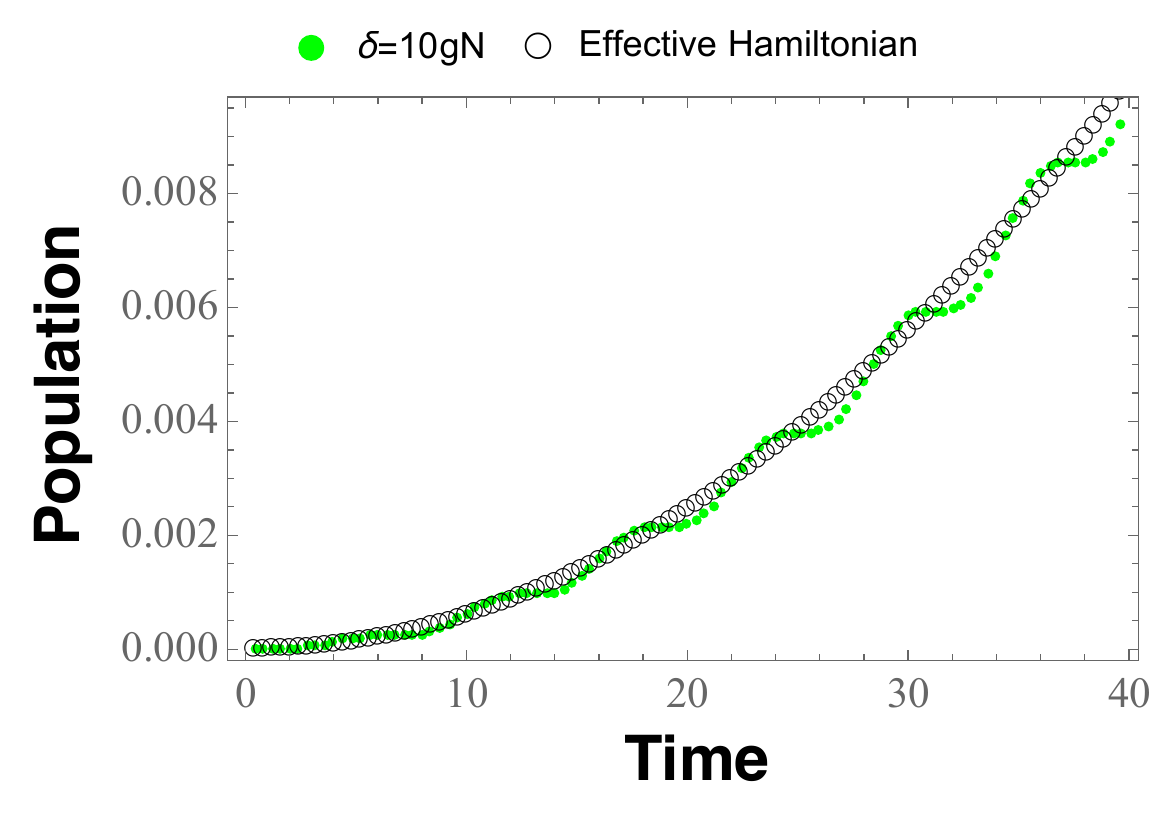}
  \end{center}
 \end{minipage}
 \begin{minipage}{0.33\hsize}
 \begin{center}
  \includegraphics[width=55mm]{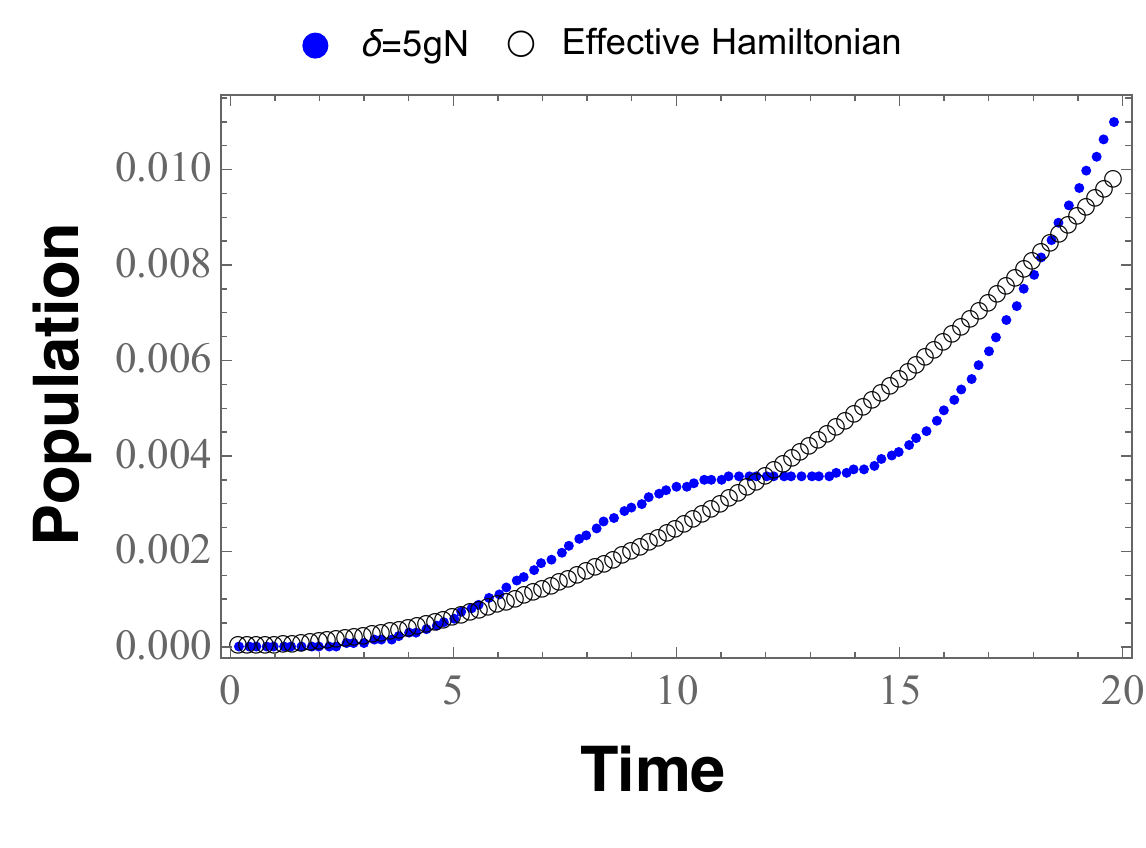}
 \end{center}
 \end{minipage}
 \begin{minipage}{0.33\hsize}
 \begin{center}
  \includegraphics[width=55mm]{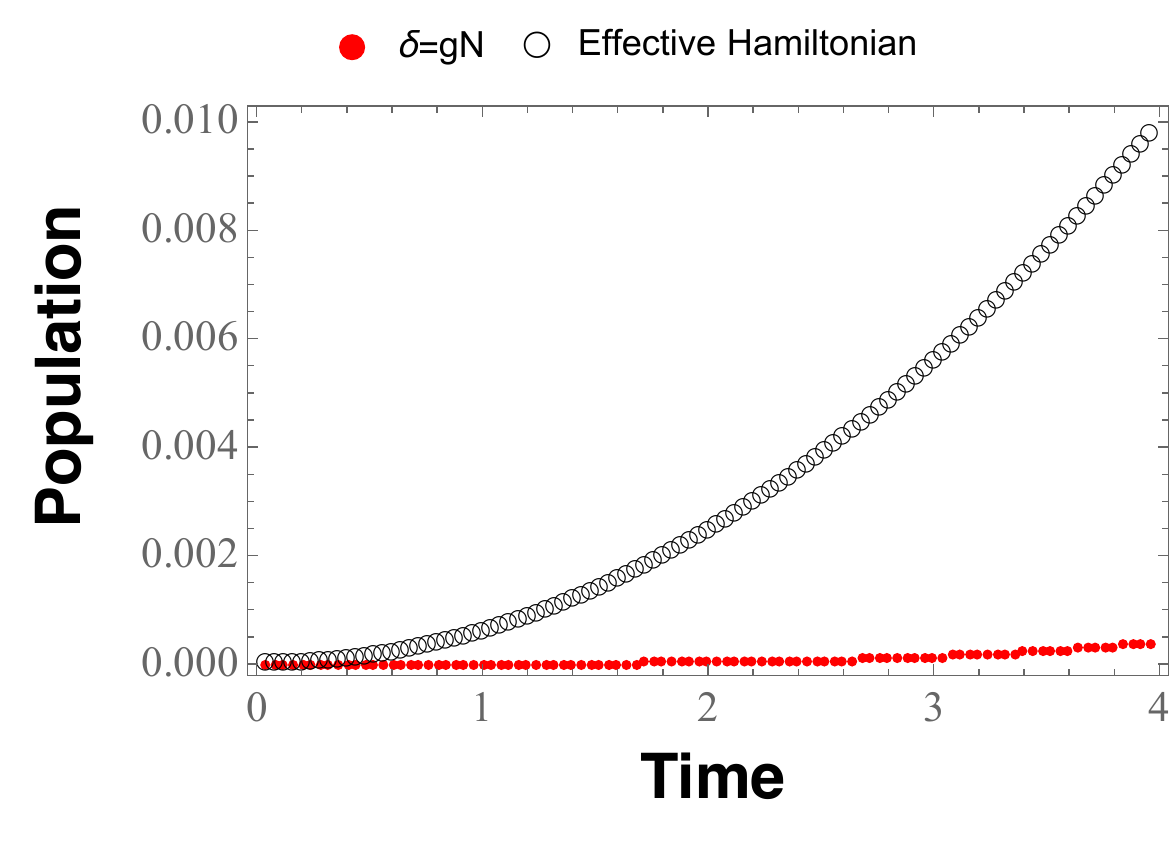}
 \end{center}
 \end{minipage}
   \caption{
   Plot of the population of $\ket{01}_B$ when we use the 
effective Hamiltonian or the exact Hamiltonian with  $\delta=10gN,\ 5gN, \ gN$.We set the other parameters as
   $g=10^{-3}, \ \omega_A=10,\ \Omega=-2.3\omega_A$ and $N=101$.}
  \label{fig:Compare_delta}
\end{figure*}

The total Hamiltonian for numerical simulations
is given by 
\begin{align}
  H_{tot}&=H_0+H_{I}\\
  H_0&=H_N+H_B
  \\
  H_I&=
  2g\ma{
  \sigma_x^{(1)}+\sigma_x^{(2)}
  }\otimes J_x\ ,
\end{align}
where we use the interaction Hamiltonian $H_I$
without the rotating wave approximation.
We consider the battery state $\rho_B(t)$ 
described as 
\begin{align}
\rho_B(t)&= 
\mathrm{Tr}_{N}\ka{
e^{-iH_{tot}t}\ket{\psi(0)}\bra{\psi(0)}
e^{iH_{tot}t}
}.
\end{align}
Firstly, we compare the dynamics with the effective Hamiltonian $H_{\rm{eff}}$ with that of the exact Hamiltonian $H_{tot}$.
Actually, as shown in FIG \ref{fig:Compare_omegaA} ,
the dynamics with the effective Hamiltonian is different from the exact Hamiltonian for $\omega _A=1$  where the condition of $\delta \ll \omega_A < \abs{\Omega}$ is violated.
On the other hand, 
in Fig \ref{fig:Compare_delta}, we observe a deviation of
the dynamics with the effective Hamiltonian from that of the exact Hamiltonian for
$\delta =gN$ where the condition of $gN\ll \delta$ is violated.

\begin{figure}[tbp]
\begin{center}
\small
\includegraphics[keepaspectratio, scale=0.5]{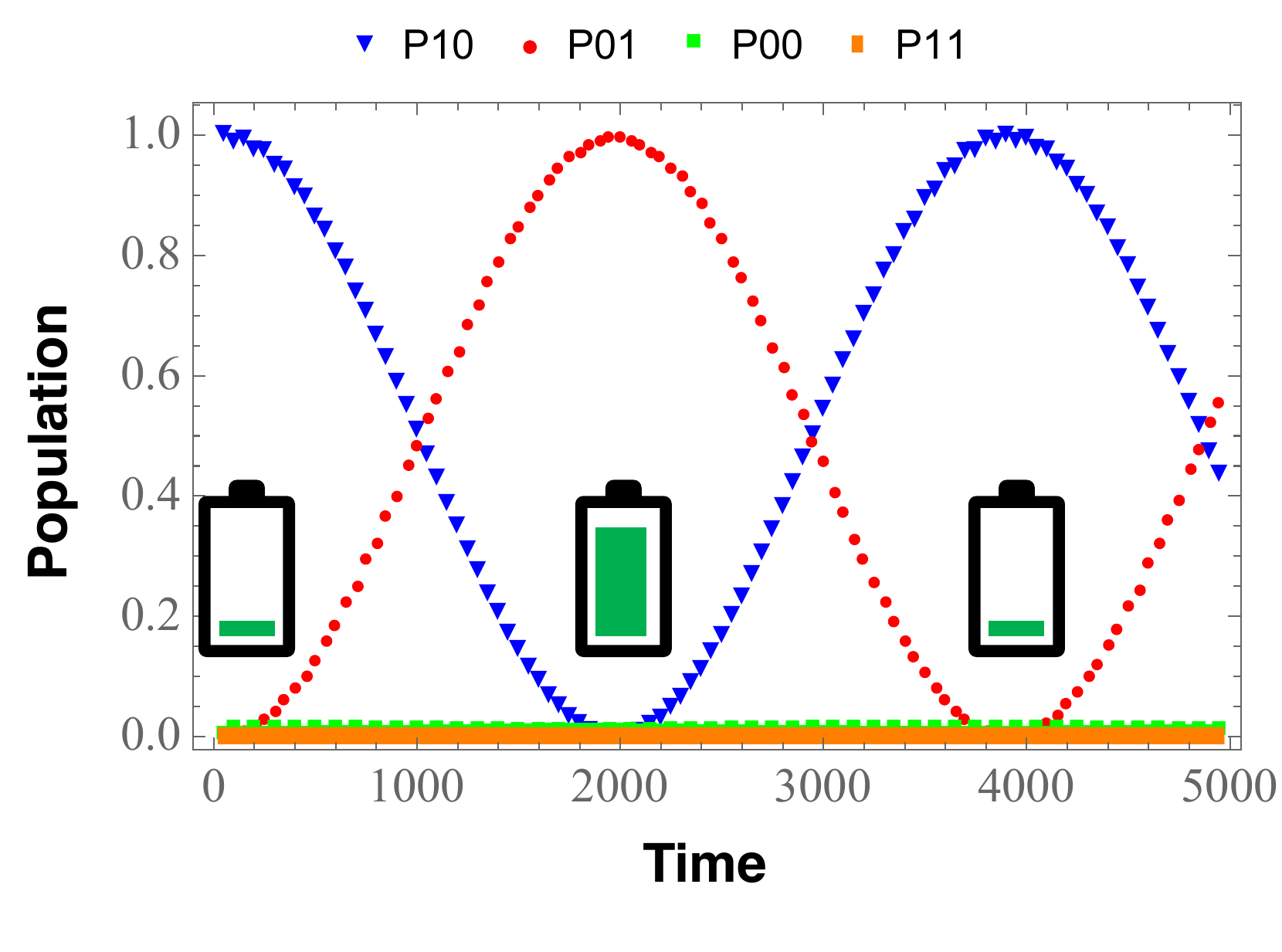}
\caption{
Plot of the population of the battery state against time.
$P_{ij}$ denotes the population of $\ket{ij}_B,\ i,j\in\na{0,1}$.
We choose the parameters as
$g=10^{-3},\ \omega_A=10,\ \delta=10gN,\ \Omega=-2.3\omega_A$ and $N=31$.
}
\label{fig:N31_population}
\end{center}
\end{figure}

\begin{figure}[tbp]
\begin{center}
\small
\includegraphics[keepaspectratio, scale=0.5]{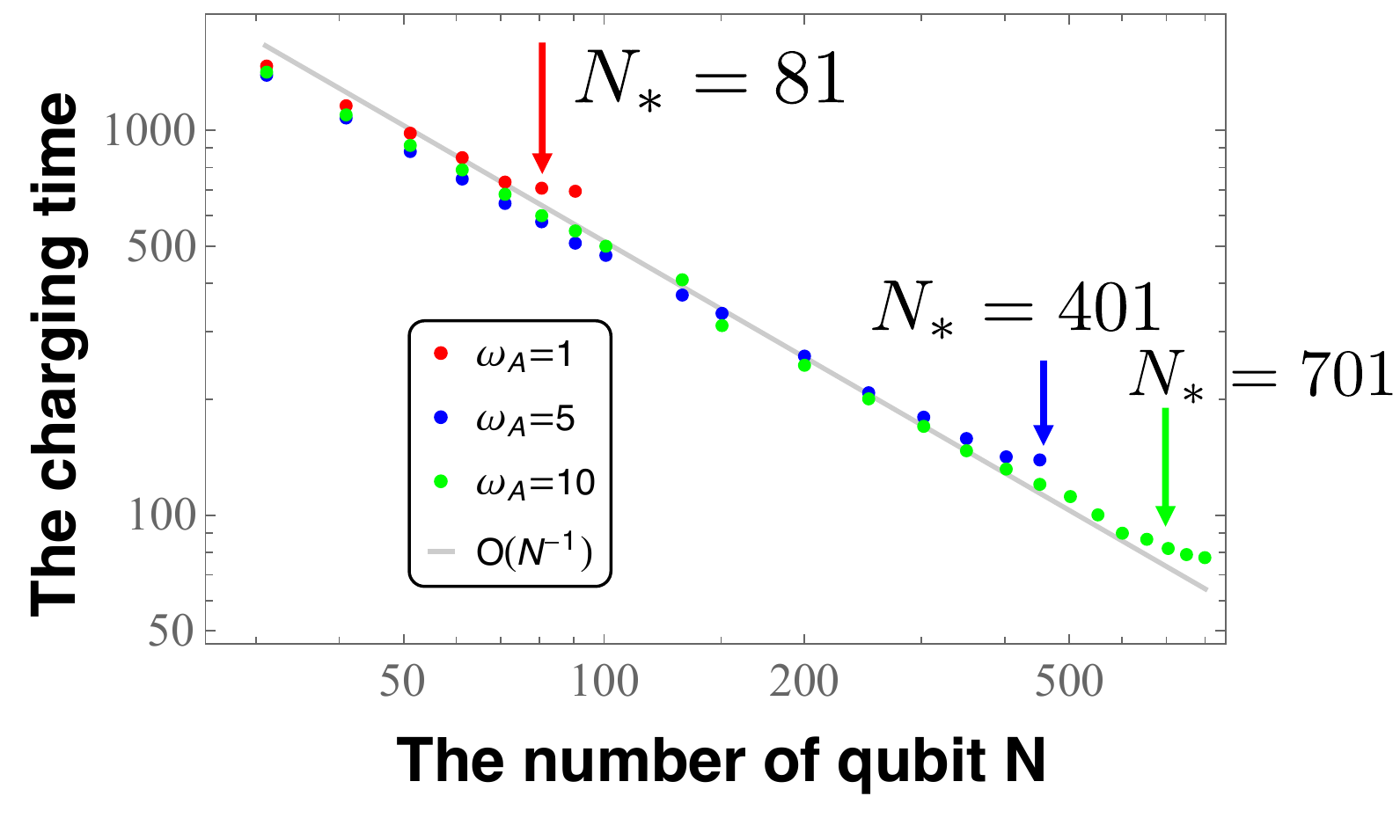}
\caption{Plot of the charging time $\bar{\tau}_N$ against the number of qubit $N$. Here, we define $N_*=\omega _A/\delta$, and our effective Hamiltonian becomes invalid when the number of the qubits becomes more than $N_*$.  We set the parameters 
as $g=10^{-3}, \ \delta=10gN$ and $\Omega=-2.3\omega_A$.
}
\label{fig:Charging_time}
\end{center}
\end{figure}

Next, we analyze the population of $\na{\ket{00}_B, \ket{01}_B, \ket{10}_B, \ket{11}_B}$ of the battery state. 
We choose the parameters to satisfy conditions of
$gN\ll \delta \ll \omega_A < \abs{\Omega}$ (See
FIG \ref{fig:N31_population}).
From the numerical results,
we confirm that there is an oscillation between $\na{\ket{10}_B}$ and $\na{\ket{01}_B}$, while a population leakage to the other states is negligible.
Secondly, we numerically calculate how the charging time depends on the number of qubit $N$ as shown in FIG \ref{fig:Charging_time}.
We introduce a ergotropy $\mathcal{E}(t)=\mathrm{Tr}_B\ka{H_B\rho_B(t)}-\min_{U:\mathrm{unitary}}\mathrm{Tr}_B\ka{H_BU\rho_B(t)U^{\dagger}}$ ~\cite{allahverdyan2004maximal},
a measure of extractable energy from battery.
We assume that the charging is done when the ergotropy of the quantum battery can be stored up to 80\% of its maximum value, and we call the necessary time for this a charging time $\bar{\tau}_N$. 
In FIG \ref{fig:Charging_time}, we 
numerically confirm that
$\bar{\tau}_N$ scale as $\mathcal{O}(N^{-1})$ in $gN\ll \delta \ll \omega_A < \abs{\Omega}$. 
However, in the region 
$\delta \simeq \omega_A
\Leftrightarrow N \simeq N_*= \frac{\omega_A}{\delta}$ where $\delta=10gN$,
we cannot finish the charging process because the target ergotropy cannot be 
80\% of its maximum value. 
This comes from the fact that the detuning $\delta$ is not strong enough to confine the dynamics into the subspace spanned by $\na{\ket{10}_B, \ket{01}_B}$. 
In other words, the first order term in the Eq. \eqref{ymadd} has a non-negligible contribution to the dynamics.

\section{Conclusion}
In conclusion, we propose a quantum battery with a charger system composed of
$N$-entangled qubits. We
ulitize a superabsoption for an entanglement-enhanced energy exchange between the charger system and quantum battery.
Our scheme provides a scaling of a charging time $\Theta\ma{N}$ while a conventional scheme
provides a scaling of a charging time $\Theta\ma{N^{-1/2}}$. 
Our results pave the way for the realization of a ultra-fast quantum battery.

\begin{acknowledgments}
This work was supported by MEXT's Leading Initiative for Excellent Young Researchers, KAKENHI (20H05661),JST PRESTO (Grant No. JPMJPR1919) and JST’s Moonshot R\&D (Grant No. JP-MJMS2061), Japan. 
\end{acknowledgments}

\appendix

\section{Derivation of the effective Hamiltonian}
In this section, we derive the effective Hamiltonian
\begin{align*}
H_{\mathrm{eff}}&=
\frac{g^2}{\delta}
  \sqrt{a_{1/2}}\sqrt{a_{3/2}}
  \ma{
  \tilde{L}_++\tilde{L}_-
  },
  \\
  \tilde{L}_+&=
  \sigma^{(2)}_+\sigma^{(1)}_-\ket{3/2}_C\bra{-1/2}
  ,\ \tilde{L}_-=\ma{\tilde{L}_+}^{\dagger}
\end{align*}
from the unitary operator (\ref{ymadd}).
We define
$\mu_1=
  \frac{1}{2}
  \ma{
  \frac{2\omega_A+\delta}{\abs{\Omega}}
  +1
  }
  $,
  $ \mu_2=
  \frac{1}{2}\ma{
  \frac{\delta}{\Omega}+1
  }$, $  \mu_3=\frac{1}{2}\ma{
  \frac{\delta}{\abs{\Omega}}+3
  }$, and $\mu_4=\frac{1}{2}\ma{
\frac{\delta-2\omega_A}{\Omega}-1
}$.
We adjust parameters to satisfy the following
\begin{eqnarray}
&&\mu_i \neq M, \ \ \nonumber \\
&&M\in \na{-N/2, -N/2+1,\cdots, N/2}
\end{eqnarray}
for all $\mu_i,\ i\in\na{1,2,3,4}$.
First, we expand the first order of interaction term
$\int^{t}_0 H^i_I(t^{\prime})dt^{\prime}$.
\begin{align}
  &\int^{t}_0 H^i_I(t^{\prime})dt^{\prime}
  \nonumber
  \\
  &=
  g\int^t_0
  \bigg\{
  \sum_{M=3/2}^{N/2}
  e^{i\ma{
  \omega_A+\delta+\Delta_M
  }t^{\prime}}
  \sqrt{a_M}\sigma^1_+\ket{M}\bra{M-1}
  \nonumber
  \\
  &
  +
  \sum_{M=-N/2}^{1/2}
  e^{i\ma{
  \omega_A+\delta-\Delta_M
  }t^{\prime}}
  \sqrt{a_M}\sigma^1_+\ket{M-1}\bra{M}
 \bigg\}
 \nonumber
  \\
  &+
  g\int^t_0
  \bigg\{
  \sum_{M=3/2}^{N/2}
  e^{i\ma{
  \abs{\omega_A+2\Omega}+\delta+\Delta_M
  }t^{\prime}}
  \sqrt{a_M}\sigma^2_+\ket{M}\bra{M-1}
  \nonumber
  \\
  &
  +
  \sum_{M=-N/2}^{1/2}
  e^{i\ma{
  \abs{\omega_A+2\Omega}+\delta-\Delta_M
  }t^{\prime}}
  \sqrt{a_M}\sigma^2_+\ket{M-1}\bra{M}
  \bigg\}
  \nonumber
  \\
  &
  +\mathrm{h.c.}
  \label{sup:process1}
\end{align}
It is worth mentioning that 
the first order of the interaction term induces transitions between states with a large energy difference. In this case, we have terms to oscillate with a high frequency, and these tend to be small.
On the other hand, these terms also have a collective enhancement factor of $\sqrt{a_M}$. We are going to evaluate whether these can be negligible or not as a total. 
\begin{align}
  \begin{split}
  &(\mathrm{\ref{sup:process1}})
    \nonumber
  \\
  &=
\sum_{M=3/2}^{N/2}
\frac{g}{i\ma{
\omega_A+\delta+\Delta_M
}}\ma{
e^{i\ma{\omega_A+\delta+\Delta_M}t}-1
}
\\
&\qquad \qquad \sqrt{a_M}\sigma^1_+\ket{M}\bra{M-1}
\nonumber
\\
&+
\sum_{M=-N/2}^{1/2}
\frac{g}{i\ma{
\omega_A+\delta-\Delta_M
}}\ma{
e^{i\ma{\omega_A+\delta-\Delta_M}t}-1
}
\\
&\qquad \qquad\sqrt{a_M}\sigma^1_+\ket{M-1}\bra{M}
\nonumber
\\
&
+
\sum_{M=3/2}^{N/2}
\frac{g}{i\ma{
\abs{\omega_A+2\Omega}+\delta+\Delta_M
}
}
\ma{
e^{i\ma{\abs{
\omega_A+2\Omega
}+\delta+\Delta_M}t}-1
}
\\
&\qquad \qquad\sqrt{a_M}\sigma_+^2\ket{M}\bra{M-1}
\nonumber
\\
&
+
\sum_{M=-N/2}^{1/2}
\frac{g}{i\ma{
\abs{\omega_A+2\Omega}+\delta-\Delta_M
}
}
\ma{
e^{i\ma{\abs{
\omega_A+2\Omega
}+\delta-\Delta_M}t}-1
}
\\
&\qquad \qquad\sqrt{a_M}\sigma_+^2\ket{M-1}\bra{M}
\nonumber
\\
&+\mathrm{h.c.}
\end{split}
\end{align}
By choosing suitable parameters, these terms are negligible as we show below.
\begin{align*}
\begin{split}
  &\frac{g}{i\ma{
  \omega_A+\delta+\Delta_M
  }}
  \sqrt{a_M}
  \\
  &\approx
  \frac{g}{2\omega_A+\delta+(2M-1)\Omega}
  \mathcal{O}(N)
    \nonumber
  \\
  &=
  \frac{g}{\delta}
  \ma{
  \frac{1}{1+(2M-1)\frac{\Omega}{\delta}
  +
  \frac{2\omega_A}{\delta}}
  }  \mathcal{O}(N)
    \nonumber
  \\
  &
  \ll 1\ \  
  \na{\because M\neq 1/2, \frac{g}{\abs{\Omega}}N\ll 1,
  \frac{\abs{\Omega}}{\delta}\gg 1
  }
  \\
  &\qquad 
  \na{\because 
  M=1/2 ,
  \frac{g}{\omega_A}N\ll 1,
  \frac{\omega_A}{\delta}\gg 1
  },
  \end{split}
  \end{align*}
\begin{align*}
\begin{split}
  &\frac{g}{i\ma{
  \omega_A+\delta-\Delta_M
  }}
  \sqrt{a_M}
  \\
  &\approx
  \frac{g}{\delta-(2M-1)\Omega}
  \mathcal{O}(N)
    \nonumber
  \\
  &=
  \frac{g}{\delta}
  \ma{
  \frac{1}{1+(2M-1)\frac{\Omega}{\delta}
  } } \mathcal{O}(N)
    \nonumber
  \\
  &\ll 1\ \ \na{\because M\neq 1/2, \frac{g}{\abs{\Omega}}N\ll 1,  \frac{\abs{\Omega}}{\delta}\gg 1}
  \\
  &\qquad 
  \na{\because M=1/2, \frac{g}{\delta}N\ll 1},
  \end{split}
\end{align*}
\begin{align*}
\begin{split}
  &\frac{g}{i\ma{
  \abs{\omega_A+2\Omega}+\delta+\Delta_M
  }}
  \sqrt{a_M}
  \\
  &\approx
  \frac{g}{\delta+(2M-3)\Omega}
  \mathcal{O}(N)
    \nonumber
  \\
  &=
  \frac{g}{\delta}
  \ma{
  \frac{1}{1+(2M-3)\frac{\Omega}{\delta}}
  }  \mathcal{O}(N)
    \nonumber
  \\
  &\ll 1\ \ \na{\because 
  M\neq 3/2,
  \frac{g}{\abs{\Omega}}N\ll 1 , \frac{\abs{\Omega}}{\delta}\gg 1
  }
  \\
  &\qquad 
  \na{\because 
  M=3/2, \frac{g}{\delta}N\ll 1
  },
  \end{split}
\end{align*}

\begin{align*}
\begin{split}
  &\frac{g}{i\ma{
  \abs{\omega_A+2\Omega}+\delta-\Delta_M
  }}
  \sqrt{a_M}
  \\
  &\approx
  \frac{g}{-2\omega_A+\delta-(2M+1)\Omega}
  \mathcal{O}(N)
    \nonumber
  \\
  &=
  \frac{g}{\delta}
  \ma{
  \frac{1}{1-(2M+1)\frac{\Omega}{\delta}
    -2\frac{\omega_A}{\delta}}
  }  \mathcal{O}(N)
    \nonumber
  \\
  &
  \ll 1\ \ 
  \na{\because M\neq-1/2,  \frac{g}{\abs{\Omega}}N\ll 1,  \frac{\abs{\Omega}}{\delta}\gg 1
  }
  \\
  &\qquad 
  \na{\because 
  M=-1/2,
  \frac{g}{\omega_A}N\ll 1,  \frac{\omega_A}{\delta}\gg 1
  }.
  \end{split}
\end{align*}
Therefore, 
we can drop the term of $\int^t_0 H_I^i(t^{\prime}) dt^{\prime}$ for $ gN \ll \delta \ll \omega_A < \abs{\Omega}.$
Next, we expand the second order of \YM{the}
interaction term 
$\int^{t}_0 H^i_I(t^{\prime})dt^{\prime}
  \int^{t^{\prime}}_0
  H^i_I(t^{\prime\prime})
  dt^{\prime\prime}$.
Since $\mu_i\neq M $ for all $i$, 
we obtain
\begin{align}
&
\int^{t}_0 H^i_I(t^{\prime})dt^{\prime}
  \int^{t^{\prime}}_0
  H^i_I(t^{\prime\prime})
  dt^{\prime\prime}
  \notag
\\
&=
\Biggl(
\sum_{1,M=3/2}^{N/2}
\frac{g^2}{i\ma{
\omega_A+\delta+\Delta_M
}}
\notag
\\
&\quad 
\int^{t}_0
dt^{\prime}
A^i(t^{\prime})
\ma{
e^{i\ma{\omega_A+\delta+\Delta_M}t^{\prime}}-1
}\sqrt{a_M}\sigma^1_+\ket{M}\bra{M-1}
  \notag
\\
&+
\sum_{2,M=-N/2}^{1/2}
\frac{g^2}{i\ma{
\omega_A+\delta-\Delta_M
}}
\notag
\\
&\quad 
\int^{t}_0
dt^{\prime}
A^i(t^{\prime})
\ma{
e^{i\ma{\omega_A+\delta-\Delta_M}t^{\prime}}-1
}\sqrt{a_M}\sigma^1_+\ket{M-1}\bra{M}
  \notag
\\
&
+
\sum_{3,M=3/2}^{N/2}
\frac{g^2}{i\ma{
\abs{\omega_A+2\Omega}+\delta+\Delta_M
}
}
\notag
\\
&\quad 
\int^{t}_0
dt^{\prime}
A^i(t^{\prime})
\ma{
e^{i\ma{\abs{
\omega_A+2\Omega
}+\delta+\Delta_M}t^{\prime}}-1
}\sqrt{a_M}\sigma_+^2\ket{M}\bra{M-1}
  \notag
\\
&+
\sum_{4,M=-N/2}^{1/2}
\frac{g^2}{i\ma{
\abs{\omega_A+2\Omega}+\delta-\Delta_M
}
}
\notag
\\
&
\int^{t}_0
dt^{\prime}
A^i(t^{\prime})
\ma{
e^{i\ma{\abs{
\omega_A+2\Omega
}+\delta-\Delta_M}t^{\prime}}-1
}\sqrt{a_M}\sigma_+^2\ket{M-1}\bra{M}
\Biggl)
\label{second_first}
\\
&-
\Biggl(
\sum_{1,M=3/2}^{N/2}
\frac{g^2}{i\ma{
\omega_A+\delta+\Delta_M
}}
\notag
\\
&\quad 
\int^{t}_0
dt^{\prime}
A^i(t^{\prime})
\ma{
e^{-i\ma{\omega_A+\delta+\Delta_M}t^{\prime}}-1
}\sqrt{a_M}\sigma^1_-\ket{M-1}\bra{M}
  \notag
\\
&+
\sum_{2,M=-N/2}^{1/2}
\frac{g^2}{i\ma{
\omega_A+\delta-\Delta_M
}}
\notag
\\
&\quad 
\int^{t}_0
dt^{\prime}
A^i(t^{\prime})
\ma{
e^{-i\ma{\omega_A+\delta-\Delta_M}t^{\prime}}-1
}\sqrt{a_M}\sigma^1_-\ket{M}\bra{M-1}
  \notag
\\
&
+
\sum_{3,M=3/2}^{N/2}
\frac{g^2}{i\ma{
\abs{\omega_A+2\Omega}+\delta+\Delta_M
}
}
\notag
\\
&\quad 
\int^{t}_0
dt^{\prime}
A^i(t^{\prime})
\ma{
e^{-i\ma{\abs{
\omega_A+2\Omega
}+\delta+\Delta_M}t^{\prime}}-1
}\nonumber
\\
&\sqrt{a_M}\sigma_-^2\ket{M-1}\bra{M}
  \notag
\\
&+
\sum_{4,M=-N/2}^{1/2}
\frac{g^2}{i\ma{
\abs{\omega_A+2\Omega}+\delta-\Delta_M
}
}
\notag
\\
&\quad 
\int^{t}_0
dt^{\prime}
A^i(t^{\prime})
\ma{
e^{-i\ma{\abs{
\omega_A+2\Omega
}+\delta-\Delta_M}t^{\prime}}-1
} \nonumber
\\
& \sqrt{a_M}\sigma_-^2\ket{M}\bra{M-1}
\Biggl)
\label{second_second}
\\
  &+\Biggl(
  \sum_{1,M=3/2}^{N/2}
  \frac{g^2}{i\ma{
  \omega_A+\delta+\Delta_M
  }}
  \nonumber
  \\
  &\quad 
  \int^{t}_0
  dt^{\prime}
  A^{i\dagger}(t^{\prime})
  \ma{
  e^{i\ma{\omega_A+\delta+\Delta_M}t^{\prime}}-1
  }\sqrt{a_M}\sigma^1_+\ket{M}\bra{M-1}
  \nonumber
  \\
  &+
  \sum_{2,M=-N/2}^{1/2}
  \frac{g^2}{i\ma{
  \omega_A+\delta-\Delta_M
  }}
  \nonumber 
  \\
  &
  \int^{t}_0
  dt^{\prime}
  A^{i\dagger}(t^{\prime})
  \ma{
  e^{i\ma{\omega_A+\delta-\Delta_M}t^{\prime}}-1
  }\sqrt{a_M}\sigma^1_+\ket{M-1}\bra{M}
  \nonumber
  \\
  &
  +
  \sum_{3,M=3/2}^{N/2}
  \frac{g^2}{i\ma{
  \abs{\omega_A+2\Omega}+\delta+\Delta_M
  }
  }
  \nonumber 
  \\
  &
  \int^{t}_0
  dt^{\prime}
  A^{i\dagger}(t^{\prime})
  \ma{
  e^{i\ma{\abs{
  \omega_A+2\Omega
  }+\delta+\Delta_M}t^{\prime}}-1
  }\sqrt{a_M}\sigma_+^2\ket{M}\bra{M-1}
  \nonumber
  \\
  &+
  \sum_{4,M=-N/2}^{1/2}
  \frac{g^2}{i\ma{
  \abs{\omega_A+2\Omega}+\delta-\Delta_M
  }
  }
  \nonumber
  \\
  &
  \int^{t}_0
  dt^{\prime}
  A^{i\dagger}(t^{\prime})
  \ma{
  e^{i\ma{\abs{
  \omega_A+2\Omega
  }+\delta-\Delta_M}t^{\prime}}-1
  }
  \nonumber
  \\
  &\qquad \sqrt{a_M}\sigma_+^2\ket{M-1}\bra{M}
  \Biggl)
  \label{second_third}
  \\
  &-
  \Biggl(
  \sum_{1,M=3/2}^{N/2}
  \frac{g^2}{i\ma{
  \omega_A+\delta+\Delta_M
  }}
  \nonumber
  \\
  &\qquad
  \int^{t}_0
  dt^{\prime}
  A^{i\dagger}(t^{\prime})
  \ma{
  e^{-i\ma{\omega_A+\delta+\Delta_M}t^{\prime}}-1
  }\sqrt{a_M}\sigma^1_-\ket{M-1}\bra{M}
  \nonumber
  \\
  &+
  \sum_{2,M=-N/2}^{1/2}
  \frac{g^2}{i\ma{
  \omega_A+\delta-\Delta_M
  }}
  \nonumber 
  \\
  &\qquad
  \int^{t}_0
  dt^{\prime}
  A^{i\dagger}(t^{\prime})
  \ma{
  e^{-i\ma{\omega_A+\delta-\Delta_M}t^{\prime}}-1
  }\sqrt{a_M}\sigma^1_-\ket{M}\bra{M-1}
  \nonumber
  \\
  &
  +
  \sum_{3,M=3/2}^{N/2}
  \frac{g^2}{i\ma{
  \abs{\omega_A+2\Omega}+\delta+\Delta_M
  }
  }
  \nonumber 
  \\
  &\qquad 
  \int^{t}_0
  dt^{\prime}
  A^{i\dagger}(t^{\prime})
  \ma{
  e^{i\ma{\abs{
  \omega_A+2\Omega
  }+\delta+\Delta_M}t^{\prime}}-1
  }\sqrt{a_M}\sigma_-^2\ket{M-1}\bra{M}
  \nonumber
  \\
  &+
  \sum_{4,M=-N/2}^{1/2}
  \frac{g^2}{i\ma{
  \abs{\omega_A+2\Omega}+\delta-\Delta_M
  }
  }
  \nonumber
  \\
  &\quad 
  \int^{t}_0
  dt^{\prime}
  A^{i\dagger}(t^{\prime})
  \ma{
  e^{-i\ma{\abs{
  \omega_A+2\Omega
  }+\delta-\Delta_M}t^{\prime}}-1
  }\sqrt{a_M}\sigma_-^2\ket{M}\bra{M-1}
  \Biggl).
  \label{second_forth}
\end{align}
We calculate the first term of the (\ref{second_first}).
\begin{equation*}
\begin{split}
      &\sum_{1,M=3/2}^{N/2}
  \frac{g^2}{i\ma{
  \omega_A+\delta+\Delta_M
  }}
  \\
  &\int^{t}_0
  dt^{\prime}
  A^i(t^{\prime})
  \ma{
  e^{i\ma{\omega_A+\delta+\Delta_M}t^{\prime}}-1
  }\sqrt{a_M}\sigma^1_+\ket{M}\bra{M-1}
  \\
  &=
  \sum_{1,M=3/2}^{N/2}
  \frac{g^2}{i\ma{
  \omega_A+\delta+\Delta_M
  }}
  \\
  &\quad
  \int^{t}_0
  dt^{\prime}
  e^{i(\abs{\omega_A+2\Omega}+\delta)t^{\prime}}
  \ma{
  e^{i\ma{\omega_A+\delta+\Delta_M}t^{\prime}}-1
  }
  e^{i\Delta_{M+1}t}
  \\
  &\qquad  \quad
  \sqrt{a_{M+1}}
  \sqrt{a_M}\sigma^2_+\sigma^1_+
  \ket{M+1}\bra{M-1}
  \\
    &=
    \sum_{M=3/2}^{N/2}
  \frac{g^2}{i\ma{
  \omega_A+\delta+\Delta_M
  }}
  \frac{1}{2i\ma{
  \Delta_M+\delta
  }}
  \\
  &\quad
  \ma{
  e^{2i\ma{\Delta_M+\delta}t}
  -1
  }
  \sqrt{a_{M+1}}
  \sqrt{a_M}\sigma^2_+\sigma^1_+
  \ket{M+1}\bra{M-1}
  \nonumber
  \\
  &\quad+
  \sum_{M=3/2}^{N/2}
  \frac{g^2}{i\ma{
  \omega_A+\delta+\Delta_M
  }}
  \frac{1}{i\ma{-\omega_A
  +\Delta_M+\delta
  }}
    \\
  &\quad
  \ma{
  e^{i\ma{-\omega_A+\Delta_M+\delta}t}
  -1
  }
  \sqrt{a_{M+1}}
  \sqrt{a_M}\sigma^2_+\sigma^1_+
  \ket{M+1}\bra{M-1}.
  \end{split}
\end{equation*}
We assume that every $M$ satisfies
$\omega_A+\delta+(2M-1)\Omega \neq 0$
and
$\delta+(2M-1)\Omega \neq 0$.
Also,
by choosing suitable parameters, the other terms are negligible as we show below.
\begin{align*}
\begin{split}
  &\frac{g^2}{\ma{
  \omega_A+\delta+\Delta_M
  }}
  \frac{1}{\ma{
  \Delta_M+\delta
  }}
  \sqrt{a_{M+1}}\sqrt{a_M}
  \\
  &\approx
  \frac{g^2\mathcal{O}(N^2)}{\delta^2}
  \frac{1}{
  1+
  \frac{3\omega_A+2(2M-1)\Omega}
  {\delta}
  +
  \frac{(2M-1)^2\Omega^2+3(2M-1)\omega_A\Omega+2\omega_A^2}
  {\delta^2}
  }
  \\
  &\ll 1\ \ \na{\because M\neq 1/2, \frac{g^2N^2}{\abs{\Omega}^2}\ll 1
, \frac{\abs{\Omega}}{\delta}\gg 1}
\\
&\qquad \na{\because 
M=1/2 ,
\frac{g^2N^2}{\omega_A^2}\ll 1
, \frac{\omega^2_A}{\delta^2}\gg 1
},
\nonumber
\\
&
\frac{g^2}{\ma{
\omega_A+\delta+\Delta_M
}}
\frac{1}{\ma{-\omega_A
+\Delta_M+\delta
}}
  \sqrt{a_{M+1}}\sqrt{a_M}
\\
&\approx
\frac{g^2\mathcal{O}(N^2)}{\delta^2}
\frac{1}{
1+
\frac{\omega_A+2(2M-1)\Omega}{\delta}
+
\frac{
(2M-1)\Omega^2+(2M-1)\omega_A\Omega-2\omega_A^2
}{\delta^2}
}
  \nonumber
\\
&\ll 1\ 
\na{\because
M\neq 1/2,
\frac{g^2N^2}{\abs{\Omega}^2}\ll 1
,\frac{\abs{\Omega}^2}{\delta^2}\gg 1
}
\\
&\qquad 
\na{\because 
M=1/2,
\frac{g^2N^2}{\omega_A^2}\ll 1
,\frac{\omega^2_A}{\delta^2}\gg 1
}.
\nonumber
\end{split}
\end{align*}
The other three terms of (\ref{second_first}) can be small for 
$gN \ll \delta \ll \omega_A <\abs{\Omega}$.
Therefore, we can ignore the contribution from (\ref{second_first}).
When an initial state is $\ket{\psi(0)}=\ket{-1/2}_C\ket{10}_B$, 
the dominant term in
(\ref{second_second}) is
\begin{align}
&\sum_{2,M=-N/2}^{1/2}
  \frac{g^2}{i\ma{
  \omega_A+\delta-\Delta_M
  }}
  \nonumber 
  \\
  &
  \int^{t}_0
  dt^{\prime}
  A^i(t^{\prime})
  \ma{
  e^{-i\ma{\omega_A+\delta-\Delta_M}t^{\prime}}-1
  }\sqrt{a_M}\sigma^1_-\ket{M}\bra{M-1}
  \label{domione}
\end{align}
and
\begin{align}
       &\sum_{3,M=3/2}^{N/2}
  \frac{g^2}{i\ma{
  \abs{\omega_A+2\Omega}+\delta+\Delta_M
  }
  }
  \nonumber 
  \\
  &
  \int^{t}_0
  dt^{\prime}
  A^i(t^{\prime})
  \ma{
  e^{-i\ma{\abs{
  \omega_A+2\Omega
  }+\delta+\Delta_M}t^{\prime}}-1
  }
  \nonumber 
  \\
  &\qquad \sqrt{a_M}\sigma_-^2\ket{M-1}\bra{M}.
  \label{domitwo}
\end{align}
We evaluate these two terms.
First, for (\ref{domione}), we obtain
 \begin{align}
 \begin{split}
&\sum_{2,M=-N/2}^{1/2}
  \frac{g^2}{i\ma{
  \omega_A+\delta-\Delta_M
  }}
  \nonumber 
  \\
  &
  \int^{t}_0
  dt^{\prime}
  A^i(t^{\prime})
  \ma{
  e^{-i\ma{\omega_A+\delta-\Delta_M}t^{\prime}}-1
  }\sqrt{a_M}\sigma^1_-\ket{M}\bra{M-1}
  \\
  &=
  \sum_{M=-N/2}^{1/2}
  \frac{g^2}{i\ma{
  \omega_A+\delta-\Delta_M
  }}
  \nonumber
  \\
  &
  \int^{t}_0
  dt^{\prime}
  e^{i\ma{
  \abs{\omega_A+2\Omega}+\delta
  }t^{\prime}}
  \ma{
  e^{-i\ma{\omega_A+\delta-\Delta_M}t^{\prime}}-1
  }
  \\
  &\qquad \quad 
  J_{+1}(t)
  \sqrt{a_M}
  \sigma^2_+\sigma^1_-\ket{M}\bra{M-1}
  \nonumber
  \\
  &=
  \sum_{M=-N/2}^{1/2}
  \sum^{N/2}_{M^{\prime}=3/2}
  \frac{g^2}{i\ma{
  \omega_A+\delta-\Delta_M
  }}
  \\
  &
  \int^{t}_0
  dt^{\prime}
  e^{i\ma{
  \abs{\omega_A+2\Omega}+\delta
  }t^{\prime}}
  \ma{
  e^{-i\ma{\omega_A+\delta-\Delta_M}t^{\prime}}-1
  }
 e^{i\Delta_{M^{\prime}}t}
 \\
 &\qquad \quad 
  \sqrt{a_M}\sqrt{a_{M^{\prime}}}
  \sigma^2_+\sigma^1_-
  \delta_{M^{\prime},M+1}
  \ket{M^{\prime}}\bra{M-1}
    \nonumber
  \\
  &=
  \frac{g^2}{i\ma{
  \omega_A+\delta-\Delta_{1/2}
  }}
  \\
  &
  \int^{t}_0
  dt^{\prime}
  e^{i\ma{
  \abs{\omega_A+2\Omega}t^{\prime}
  }}
  \ma{
  e^{-i\ma{\omega_A+\delta-\Delta_{1/2}}t^{\prime}}-1
  }
 e^{i\Delta_{3/2}t}
 \\
 &\qquad \quad 
  \sqrt{a_{1/2}}\sqrt{a_{3/2}}
  \sigma^2_+\sigma^1_-
  \ket{3/2}\bra{-1/2}
  \nonumber
  \\
  &=
  \frac{g^2}{i\ma{
  \omega_A+\delta-\omega_A
  }}
  \\
  &
  \int^{t}_0
  dt^{\prime}
  e^{i\ma{
  \abs{\omega_A+2\Omega}+\delta
  }t^{\prime}}
  \ma{
  e^{-i\ma{\omega_A+\delta-\omega_A}t^{\prime}}-1
  }
 e^{i\ma{\omega_A+2\Omega}t}
 \\
 &\qquad \quad 
  \sqrt{a_{1/2}}\sqrt{a_{3/2}}
  \sigma^2_+\sigma^1_-
  \ket{3/2}\bra{-1/2}
  \nonumber
  \\
  &=
  \frac{g^2}{i\delta}
  \int^{t}_0
  dt^{\prime}
  \ma{
  1-e^{i\delta t^{\prime}}
  }
  \sqrt{a_{1/2}}\sqrt{a_{3/2}}
  \sigma^2_+\sigma^1_-
  \ket{3/2}\bra{-1/2}
  \nonumber
  \\
  &=
  \frac{g^2}{i\delta}
  \ma{
  t-\frac{1}{i\delta}\ma{
  e^{i\delta t}-1
  }
  }
  \sqrt{a_{1/2}}\sqrt{a_{3/2}}
  \sigma^2_+\sigma^1_-
  \ket{3/2}\bra{-1/2}
    \nonumber
  \\
  &\approx
  -\frac{ig^2t}{\delta}
  \sqrt{a_{1/2}}\sqrt{a_{3/2}}
  \sigma^2_+\sigma^1_-
  \ket{3/2}\bra{-1/2},
  \ \ 
  \end{split}
 \end{align}
where we use $1\ll \delta t$ in the last line.
 Also, we impose 
 a condition of
 $1\ll gNt$ so
 that the second order $\frac{(gN)^2t}{\delta}$ of the interaction
should be
 larger than the first order $\frac{gN}{\delta}$.
 Since we assume
 $1\ll gNt$  and $gN\ll \delta$, 
 the necessary condition is written as
 $1\ll gNt \ll \delta t$. 
Second, for (\ref{domitwo}), we obtain
 \begin{align*}
   &\sum_{3,M=3/2}^{N/2}
  \frac{g^2}{i\ma{
  \abs{\omega_A+2\Omega}+\delta+\Delta_M
  }
  }
  \nonumber 
  \\
  &
  \int^{t}_0
  dt^{\prime}
  A^i(t^{\prime})
  \ma{
  e^{-i\ma{\abs{
  \omega_A+2\Omega
  }+\delta+\Delta_M}t^{\prime}}-1
  }\sqrt{a_M}\sigma_-^2\ket{M-1}\bra{M}
  \\
  &\approx
  -i\frac{g^2t}{\delta}
  \sqrt{a_{1/2}}
  \sqrt{a_{3/2}}
  \sigma_+^1
  \sigma_-^2
  \ket{-1/2}\bra{3/2}
  \ \
  \ma{
  1\ll \delta t
  }.
 \end{align*}
From these calculations,
 the second order of the
 interaction can be described as 
 \begin{align*}
  \int^{t}_0 H^i_I(t^{\prime})dt^{\prime}
  \int^{t^{\prime}}_0
  H^i_I(t^{\prime\prime})
  dt^{\prime\prime}
  &\approx 
  \frac{ig^2t}{\delta}
  \sqrt{a_{1/2}}\sqrt{a_{3/2}}
  \ma{
  \tilde{L}_++\tilde{L}_-
  }
  \\
  &
  \ \ \ma{
  \tilde{L}_+\equiv \sigma^2_+\sigma^1_-\ket{3/2}\bra{-1/2}
  }.
\end{align*}
Therefore, we obtain the following.
\begin{eqnarray}
|\psi(t)\rangle 
\simeq |\psi(0)\rangle 
  -\frac{ig^2t}{\delta}
  \sqrt{a_{1/2}}\sqrt{a_{3/2}}
  \ma{
  \tilde{L}_++\tilde{L}_-
  }
\label{ymadd}.
\end{eqnarray}
We can this rewrite as 
\begin{eqnarray}
\frac{|\psi(t)\rangle  
-|\psi(0)\rangle 
}{t}=\frac{ig^2t}{\delta}
  \sqrt{a_{1/2}}\sqrt{a_{3/2}}
  \ma{
  \tilde{L}_++\tilde{L}_-
  }\nonumber
\end{eqnarray}
and, we obtain
\begin{eqnarray}
\frac{d|\psi (t)\rangle
}{dt}\simeq -iH_{\mathrm{eff}} |\psi (t)\rangle \label{ymsh}
\end{eqnarray}
where we define the effective Hamiltonian as 
\begin{align*}
    H_{\mathrm{eff}}\equiv 
     \frac{ig^2t}{\delta}
  \sqrt{a_{1/2}}\sqrt{a_{3/2}}
  \ma{
  \tilde{L}_++\tilde{L}_-
  }.
\end{align*}
By solving the (\ref{ymsh}), we obtain a state 
at a time $t$ with the effective Hamiltonian.
This kind of approximation has been used
in quantum optics \cite{gerry2005introductory}, but we firstly apply this technique to the
model of the quantum battery with a superabsorption.

~~~


\bibliography{kamimura}
\end{document}